\documentclass[twocolumn]{aastex62}
\graphicspath{{./}{figures/}}

\received{July 23, 2020}
\revised{August 1, 2020}
\accepted{August 10, 2020}
\submitjournal{ApJ}

\shortauthors{Li et al.}

\begin{document}

\title{Evidence for dense gas heated by the explosion in Orion\,KL}

\author[0000-0002-0786-7307]{Dalei Li}
\affil{Xinjiang Astronomical Observatory, Chinese Academy of Sciences, 830011 Urumqi, PR China}
\affil{Key Laboratory of Radio Astronomy, Chinese Academy of Sciences, 830011 Urumqi, PR China}

\author[0000-0002-4154-4309]{Xindi Tang}
\affil{Xinjiang Astronomical Observatory, Chinese Academy of Sciences, 830011 Urumqi, PR China}
\affil{Key Laboratory of Radio Astronomy, Chinese Academy of Sciences, 830011 Urumqi, PR China}

\author[0000-0002-7495-4005]{Christian Henkel}
\affil{Max-Planck-Institut f\"{u}r Radioastronomie, Auf dem H\"{u}gel 69, 53121 Bonn, Germany}
\affil{Astronomy Department, King Abdulaziz University, PO Box 80203, 21589 Jeddah, Saudi Arabia}
\affil{Xinjiang Astronomical Observatory, Chinese Academy of Sciences, 830011 Urumqi, PR China}

 \author[0000-0001-6459-0669]{Karl M. Menten}
 \affil{Max-Planck-Institut f\"{u}r Radioastronomie, Auf dem H\"{u}gel 69, 53121 Bonn, Germany}

 \author[0000-0003-4516-3981]{Friedrich Wyrowski}
 \affil{Max-Planck-Institut f\"{u}r Radioastronomie, Auf dem H\"{u}gel 69, 53121 Bonn, Germany}

 \author{Yan Gong}
 \affil{Max-Planck-Institut f\"{u}r Radioastronomie, Auf dem H\"{u}gel 69, 53121 Bonn, Germany}

 \author{Gang Wu}
 \affil{Xinjiang Astronomical Observatory, Chinese Academy of Sciences, 830011 Urumqi, PR China}
 \affil{Key Laboratory of Radio Astronomy, Chinese Academy of Sciences, 830011 Urumqi, PR China}

 \author[0000-0002-8760-8988]{Yuxin He}
 \affil{Xinjiang Astronomical Observatory, Chinese Academy of Sciences, 830011 Urumqi, PR China}
 \affil{Key Laboratory of Radio Astronomy, Chinese Academy of Sciences, 830011 Urumqi, PR China}

 \author{Jarken Esimbek}
 \affil{Xinjiang Astronomical Observatory, Chinese Academy of Sciences, 830011 Urumqi, PR China}
 \affil{Key Laboratory of Radio Astronomy, Chinese Academy of Sciences, 830011 Urumqi, PR China}

 \author[0000-0003-0356-818X]{Jianjun Zhou}
 \affil{Xinjiang Astronomical Observatory, Chinese Academy of Sciences, 830011 Urumqi, PR China}
 \affil{Key Laboratory of Radio Astronomy, Chinese Academy of Sciences, 830011 Urumqi, PR China}

\correspondingauthor{Xindi Tang}
\email{tangxindi@xao.ac.cn}

\begin{abstract}
We mapped the kinetic temperature structure of Orion\,KL in a $\sim$20$''$ ($\sim$8000\,AU)
sized region with para-H$_{2}$CS\,$7_{07}-6_{06}$, $7_{26}-6_{25}$, and $7_{25}-6_{24}$
making use of ALMA Band 6 Science Verification data. The kinetic temperatures obtained with
a resolution of 1\hbox{$\,.\!\!^{\prime\prime}$}65$\times$1\hbox{$\,.\!\!^{\prime\prime}$}14\,($\sim$550\,AU)
are deduced by modeling the measured averaged velocity-integrated
intensity ratios of para-H$_2$CS\,$7_{26}-6_{25}/7_{07}-6_{06}$ and $7_{25}-6_{24}/7_{07}-6_{06}$
with a RADEX non-LTE model. The kinetic temperatures of the dense gas, derived from the para-H$_2$CS
line ratios at a spatial density of 10$^7$\,cm$^{-3}$,  are high, ranging from 43 to $>$500\,K
with an unweighted average of $\sim$170\,K. There is no evidence for internal sources playing an important
role in the heating of the various structures identified in previous work, namely the elongated ridge,
the northwestern clump, and the eastern region of the
compact ridge, while the high temperatures in the western region of the compact ridge
may be dominated by internal massive star formation. Significant gradients of kinetic
temperature along molecular filaments traced by H$_2$CS indicate that the dense gas
is heated by the shocks induced by the enigmatic explosive event, which occurred several
hundred years ago greatly affecting the energetics of the Orion\,KL region.
Thus, with the notable exception of the western region of the compact ridge, the high
temperatures of the dense gas in Orion\,KL are probably caused by shocks from the
explosive event, leading to a dominant component of externally heated dense gas.
\end{abstract}

\keywords{star: formation  – star: massive  – ISM: clouds – ISM: molecules  – techniques: interferometric}

\section{Introduction}
\label{intro}
\subsection{The Energetics of the Orion KL Nebula}
The Kleinmann-Low (KL) nebula \citep{Kleinmann1967} is one of the richest known sources of molecular
line emission at millimeter and submillimeter (submm) wavelengths \citep[e.g., ][]{Blake1987}.
There are several reasons for this: First, at a distance of $\approx 400$\,pc \citep{Menten2007,Kim2008,Kounkel2017}
it is the nearest interstellar source containing a molecular spectrum that is characterized
by a high temperature, which in some parts exceeds 150\,K. Second, it is located in the most (submm) luminous
part of a 7\,pc long high column density filament \citep{Johnstone1999}
that is heated to this high temperature.  This causes a plethora of molecules to evaporate off dust grain mantles,
where they attained substantial abundances in ice form while the region was at a lower temperature.
This caused the region to become the eponymous ``hot core'' \citep{HO1979}, although, in contrast to the
many hot cores found in high-mass star forming regions it does not contain a central heating source \citep{Zapata2011}.
The energetics of the hot and dense molecular gas on a scale of a few thousand au, the topic of this study,
has been the subject of abundant previous work.

An obvious cause for the high temperatures is the energetic explosion that is thought to have occurred
in the Orion\,KL nebula about 550\,yr ago \citep{Bally2005,Zapata2009,Bally2011,Bally2017}.
Spectacular evidence comes from near-infrared (NIR) emission from shock-excited molecular hydrogen, H$_2$,
first detected by \citet{Beckwith1978} and from NIR and optical lines from neutral atoms and ions \citep{Allen1993}
that were found to form finger-like filaments \citep[e.g.,][]{Kaifu2000}. The finger-like filaments were
also detected in mm-wavelength emission from CO \citep{Zapata2009}. Emission in the CO\,$J$\,=\,2--1 line
was recently imaged with the Atacama Large Millimeter/submillimeter Array (ALMA). It shows at velocities outside of
$v_{\rm LSR}$ = 0 to 20 km\,s$^{-1}$ a large number of radial streamers covering the entire  range of position
angles and occupying a region with radius 50$^{\prime\prime}$ \citep{Bally2017}. This differs from the
NIR H$_2$ emission, part of which is absorbed by massive dust emission associated with the KL nebula \citep{Johnstone1999}.
The CO filaments and H$_2$ fingers can be traced back to a common origin
\citep{Zapata2009,Bally2011,Bally2015,Bally2017} that is consistent with the position to which the proper
motions of various radio and NIR emitting young stellar projects can be traced back in space and time \citep{Rodr2005,Gomez2005,Gomez2008,Luhman2017,Rodriguez2017,Bally2020,Rodriguez2020}.
Based on their Very Large Array (VLA) proper motion determinations, \citet{Rodriguez2017} determine the year $1445\pm6$
as the time when the most prominent radio sources, Source I (in the following Src\,I) and
the Becklin-Neugebauser object (BN), were closest together.

These findings constitute persuasive evidence the explosion resulted from the merger of young stellar objects,
at least one of them massive (Src\,I),  a scenario first discussed by \citet{Bally2005}
in which the $10^{48}$--$10^{49}$ ergs injected into the region represent binding energy released
by the orbital decay of components of a multiple system. More detailed observations
suggest a more complex picture, a.o., suggesting ejections of sources at later times \citep{Rodriguez2020}.

\begin{figure*}[t]
\centering
\epsscale{1.20}
\plotone{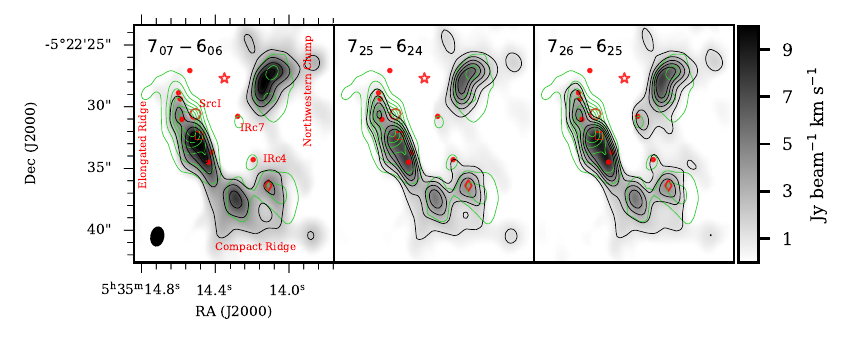}
\caption{Gray maps and black contours show velocity-integrated intensity distributions of
para-H$_2$CS $7_{07}-6_{06}$ (\emph{left}), $7_{25}-6_{24}$ (\emph{middle}),
and $7_{26}-6_{25}$ (\emph{right}). Velocities were  integrated  from
$V_{\rm LSR}$\,=\,3 to 11\,km\,s$^{-1}$. The black contour levels start at
2.2\,Jy\,beam$^{-1}$\,km\,s$^{-1}$ ($\sim$32$\sigma$) and increase in intervals of
2.2\,Jy\,beam$^{-1}$\,km\,s$^{-1}$ for para-H$_2$CS $7_{07}-6_{06}$ (\emph{left}) and at
1.1\,Jy\,beam$^{-1}$\,km\,s$^{-1}$ ($\sim$16$\sigma$) with spacings  of
1.1\,Jy\,beam$^{-1}$\,km\,s$^{-1}$ for para-H$_2$CS $7_{25}-6_{24}$
(\emph{middle}) and $7_{26}-6_{25}$ (\emph{right}), respectively.
The green contours represent the 1.2\,mm continuum emission
with levels from 0.07 ($\sim$7$\sigma$) to 1.07\,Jy\,beam$^{-1}$ and steps of 0.2\,Jy\,beam$^{-1}$.
Red star, red circle, red square, red diamond, and red points show the locations
of the putative explosive center, Src\,I, SMA1, H$_2$O maser burst,
and compact continuum sources \citep{Hirota2015},
respectively. The beam size is shown in the lower left corner.}
\label{fig1}
\end{figure*}

The radio source Src\,I \citep{Churchwell1987,Menten1995}, the primary remnant source of the erstwhile
multiple stellar system, has been speculated to be a close binary itself that is
surrounded by a Keplerian disk whose rotation curve indicates a mass of 15\,M$_{\odot}$ \citep{Ginsburg2018DISK}.
To make an already complex situation even more complicated,
Src\,I is driving a bipolar molecular outflow along a northeast-southwest axis that may also
contribute to the KL region's energetics \citep{Plambeck2009OUTFLOW}.

Indeed, imaging of the high excitation NH$_3$\,$(J,K)\,=\,(6,6)-(12,12)$ inversion transitions suggests
that substantial parts of the NH$_3$ bearing hot molecuclar gas in Orion\,KL may be heated by the outflow from
Src\,I and also by ram pressure due to the source's motion reaching NH$_3$ temperatures up to $\sim$490\,K \citep{Goddi2011}.
This picture is corroborated by observations of CH$_3$CN\,$J$\,=\,18--17 reported by \citet{Wang2010}.
Within this context, \citet{Favre2011} and \citet{Wang2011} suggested that this outflowing
gas may also heat a part of the compact ridge, which is located $\sim$8$^{\prime\prime}$ ($\sim$3300\,AU)
to the southwest (see Fig.\,\ref{fig1}).

In contrast (or complementary), \cite{Zapata2011} and \cite{Orozco2017}
proposed that the Hot Core  in the elongated ridge is externally heated by the above mentioned explosion.
Recent observations of H$_2$CO\,$J$\,=\,3--2 also indicate that the dense gas may be influenced
by this explosion \citep{Tang2018a}.

While an impact from the violent explosion on the surrounding gas in the
Orion\,KL region may be expected, it is not yet clear whether this can be considered as the dominant
heating source in at least some parts of the Orion\,KL region.

\subsection{H$_2$CS as a temperature probe}
Thioformaldehyde (H$_2$CS), like H$_2$CO a slightly asymmetric rotor molecule, exhibits a
large number of millimeter and submillimeter transitions
(e.g., \citealt{Wooten2009,Tercero2010,Widicus2017,Luo2019,Brinkmann2020,Van2020,Johnston2020}).
It is a sensitive tracer to probe the kinetic temperature, $T_{\rm kin}$, of dense gas
with line ratios involving different $K_{\rm a}$ ladders because  the relative
populations of the $K_{\rm a}$ ladders of H$_2$CS are governed by collisions
\citep{Mangum1993,Wooten2009,Oya2016,Van2020}. Since the frequencies are close for the
same $J$ transitions belonging to  different $K_{\rm a}$ ladders
(e.g., $J$\,=\,7--6, 8--7, 10--9), these transitions can be easily covered
within a single spectral band and be observed simultaneously. This  minimizes
observational uncertainties, such as differences in spatial resolution, uncertainties in absolute
flux calibration, and deviating pointing errors. Previous observations indicate that H$_2$CS
has a lower abundance than H$_2$CO in star formation regions (e.g., \citealt{Tercero2010,Nagy2015}).
The transitions of H$_2$CS at millimeter or submillimeter wavelengths  are only weakly
affected by  opacity effects (e.g., \citealt{Nagy2015,Luo2019}, and our Sect.\,\ref{H2CS_thermometer}).
Therefore, H$_2$CS can trace physical conditions of denser gas than H$_2$CO (e.g.,
\citealt{Mangum1993,Wooten2009,Tang2017a,Tang2017b,Tang2018a,Tang2018b,Van2020}).

\begin{figure*}[t]
\centering
\epsscale{1.22}
\plotone{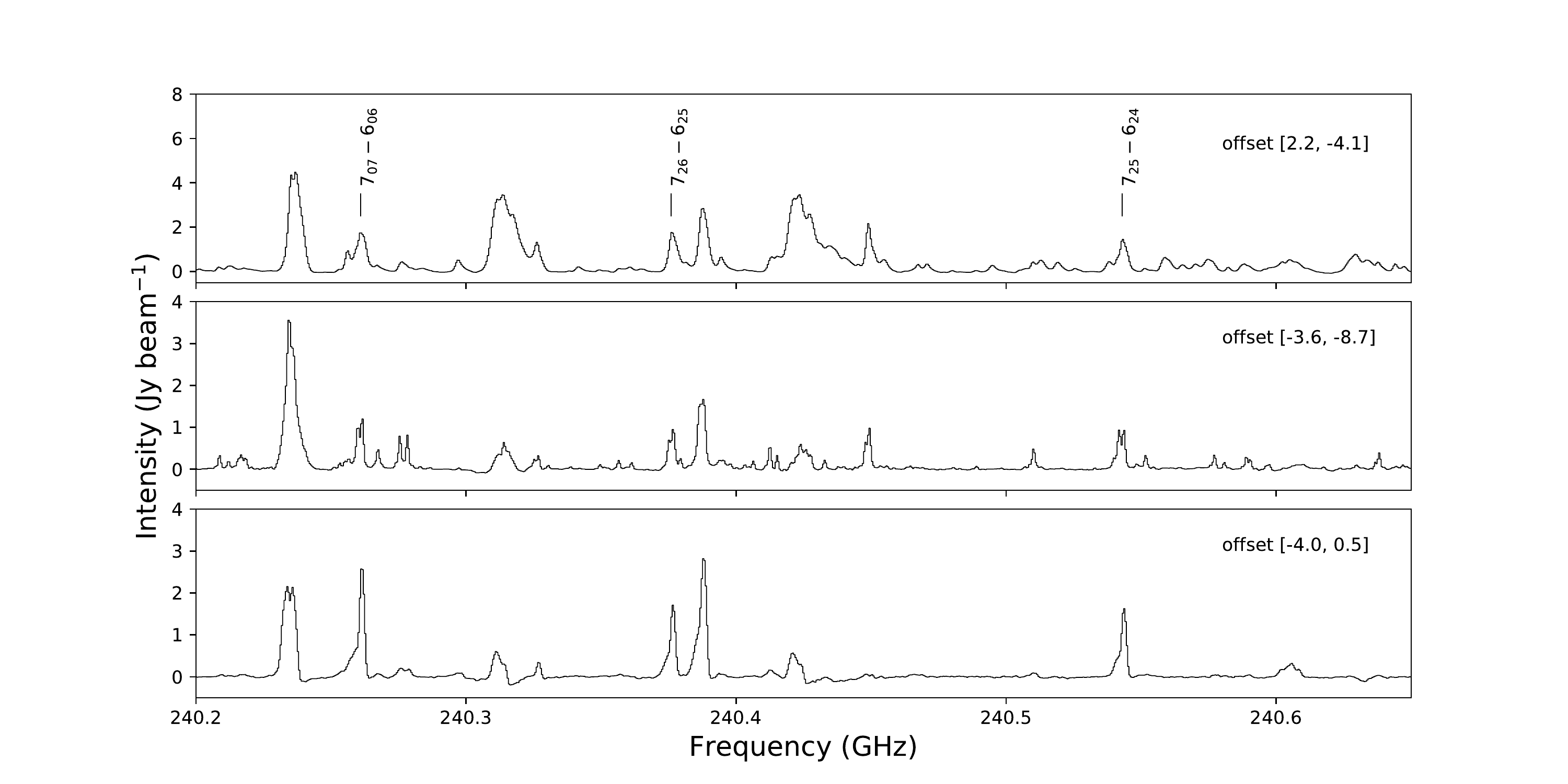}
\caption{Observed para-H$_2$CS\,$7_{07}-6_{06}$, $7_{25}-6_{24}$, and $7_{26}-6_{25}$ spectra
towards the elongated ridge (\emph{top}), the compact ridge (\emph{middle}), and the northwestern
clump (\emph{bottom}). For the morphology and nomenclature, see Fig.\,\ref{fig1}.
Each spectrum was extracted from a single pixel of size
0\hbox{$\,.\!\!^{\prime\prime}$}2$\times$0\hbox{$\,.\!\!^{\prime\prime}$}2,
but covers the entire synthesized beam size of
1\hbox{$\,.\!\!^{\prime\prime}$}65$\times$1\hbox{$\,.\!\!^{\prime\prime}$}14 (see Sect.\,\ref{Data}).
The offsets (in arcsec) are relative to the explosive center
($\alpha_{\rm 2000}$=05$^h$35$^m$14.35$^s$, $\delta_{\rm 2000}$=--05$\degr$22$'$27.7$''$)
estimated from proper motion measurements \citep{Rodr2005,Gomez2005,Gomez2008}.}
\label{fig-spect}
\end{figure*}

Making use of the attractive properties of  thioformaldehyde, we provide maps of the kinematic temperature
structure of the Orion\,KL region based on the three transitions of
para-H$_{2}$CS\,$J_{K_{\rm a}K_{\rm c}}\,=\,7_{07}-6_{06}$, $7_{26}-6_{25}$, and $7_{25}-6_{24}$,
and investigate the potential heating sources of the dense gas.
In Sects.\,\ref{Data} and \ref{results}, we introduce archival H$_2$CS data,
describe the data reduction, and outline the main observational results. The obtained
dense gas temperatures are  then discussed in Sect.\,\ref{Discussion}. Our main
conclusions are summarized in Sect.\,\ref{Conclusions}.

\section{Archival data and their analysis}
\label{Data}
The data of the para-H$_2$CS triplet were taken as part of the ALMA Science Verification.
The observations of Orion\,KL (Project: 2011.0.00009.SV)  were carried out by
ALMA in band 6 with the 12-m array including 16 antennas on 2012 January 20.
The projected length of the  baselines ranged  from 17 to 265\,m. Flux and
bandpass fluctuations were calibrated  by observations of Callisto, while the
phase fluctuations were calibrated using the  quasar J0607-085.
Twenty spectral windows  of 1.875\,GHz bandwidth and 3840 channels each were
used during the observations resulting in a spectral resolution of 0.488\,MHz
($\sim$0.61\,km\,s$^{-1}$ at 240\,GHz). Here, only the spectral
window 'spw\,12' (239.848--241.723\,GHz) was selected,  which includes all the
para-H$_2$CS\,$J$\,=\,7--6 lines used in this work.

The data reduction was performed using CASA\footnote{https://casa.nrao.edu} \citep{McMullin2007CASA}.
The calibrated data was imaged using the  \texttt{tclean} algorithm with
Briggs robust parameter 0.5, achieving a beam size of
1\hbox{$\,.\!\!^{\prime\prime}$}65$\times$1\hbox{$\,.\!\!^{\prime\prime}$}14
and setting a pixel size of 0\hbox{$\,.\!\!^{\prime\prime}$}2$\times$0\hbox{$\,.\!\!^{\prime\prime}$}2.
The cleaned image was corrected for the primary beam response of the ALMA antenna. From the corrected image,
a continuum image and a continuum subtracted line  cube were simultaneously produced
using the python-based tool STATCONT\footnote{https://hera.ph1.uni-koeln.de/$\sim$sanchez/statcont}
\citep{Sanchez2018Statcont}. The line cubes of para-H$_2$CS\,$7_{07}-6_{06}$\,(240.267\,GHz; $E_{\rm u}/k = 46.1$\,K),
$7_{25}-6_{24}$\,(240.549\,GHz; $E_{\rm u}/k = 98.8$\,K), and $7_{26}-6_{25}$\,(240.382\,GHz; $E_{\rm u}/k = 98.8$\,K)
were then extracted from the continuum subtracted line cube. The typical rms for the lines is
$\sim$30\,mJy\,beam$^{-1}$  and $\sim$10\,mJy\,beam$^{-1}$ for the continuum. Observed spectra toward
three selected positions of Orion\,KL are shown in Fig.\,\ref{fig-spect}.

\begin{figure*}[t]
\centering
\epsscale{1.12}
\plotone{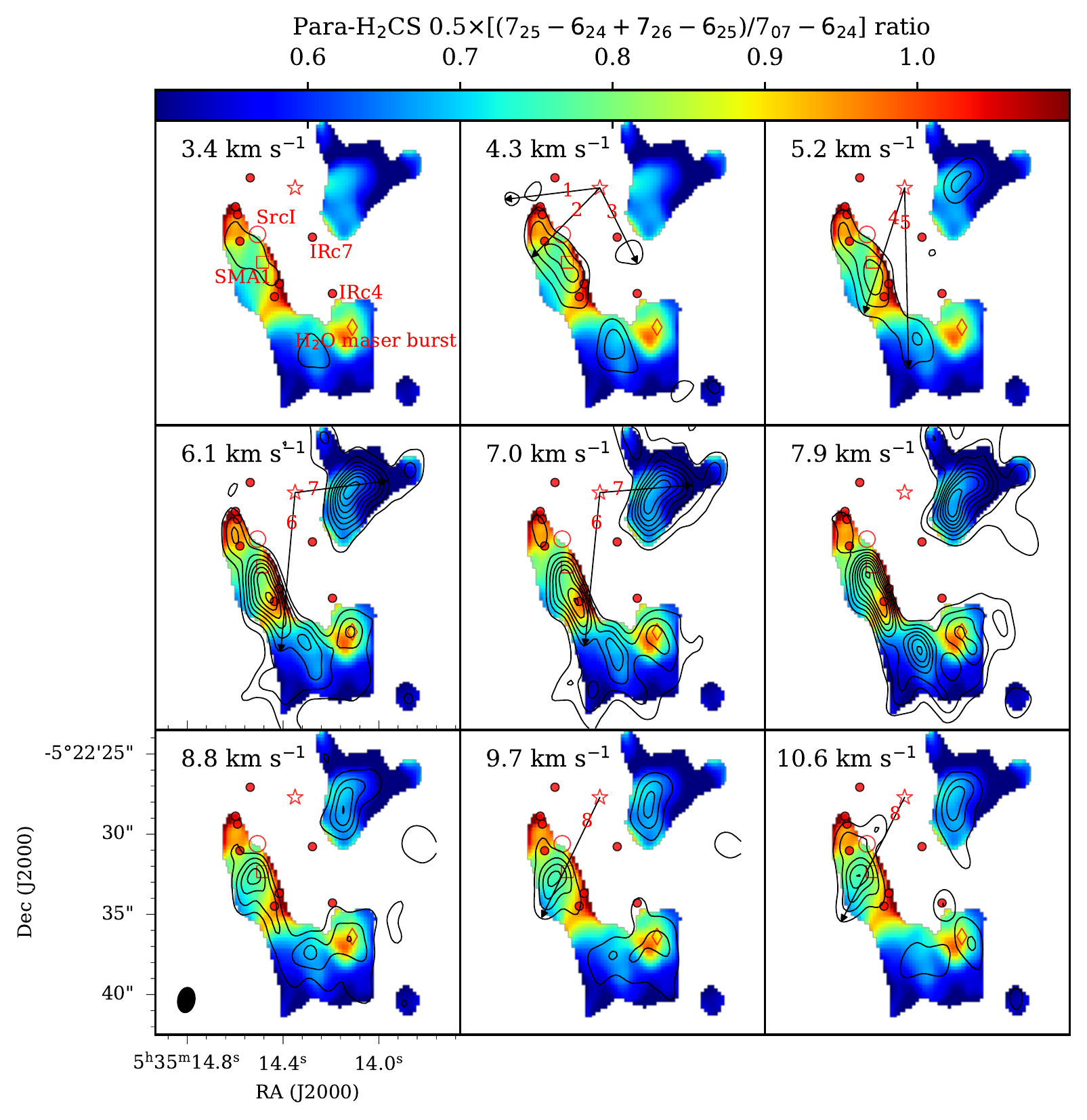}
\caption{Channel maps of para-H$_2$CS\,$7_{07}-6_{06}$ (\emph{black contours}) overlaid
on averaged (3--11\,km\,s$^{-1}$)  velocity-integrated intensity ratio maps of
para-H$_2$CS\,0.5$\times[(7_{25}-6_{24}+7_{26}-6_{25})/7_{07}-6_{06}$] (\emph{colour maps}).
The contour levels range from 10\% (0.4\,Jy\,beam$^{-1}$\,km\,s$^{-1}$) to
90\% (3.7\,Jy\,beam$^{-1}$\,km\,s$^{-1}$) of the maximum value of these channel maps.
Black arrows, starting from the location of the explosive event \citep{Pagani2019III},
show  directions of the potential displacements of gas due to this event.
Red star, red circle, red square, red diamond, and red points are the same as in Fig.\,\ref{fig1}.
The pixel size of each image is 0\hbox{$\,.\!\!^{\prime\prime}$}2$\times$0\hbox{$\,.\!\!^{\prime\prime}$}2.
The beam size is shown in the lower left corner.}
\label{Fig2}
\end{figure*}

\section{Results}
\label{results}
\subsection{Continuum emission}
The $\sim$1.2\,mm\,(240\,GHz) continuum emission distribution is presented in Fig.\,\ref{fig1}.
It  is mainly associated with the elongated ridge seen on large scale \citep{Johnstone1999}
with the major axis along northeast-southwest direction,
the compact ridge and the northwestern clump of the Orion\,KL region. This is consistent with previous
observational results obtained with the Submillimeter Array  (SMA; e.g., \citealt{Tang2010,Feng2015SMA}),
Plateau de Bure Interferometer (PdBI; e.g., \citealt{Favre2011}), and ALMA
(e.g., \citealt{Hirota2015,Pagani2017I}). In the Orion\,KL region,
eleven compact sources (see Fig.\,\ref{fig1}) with convolved
sizes $\textless1^{\prime\prime}$ ($\textless$450\,AU) and  volume densities
$10^{8}-10^{9}$\,cm$^{-3}$, including well known sources such as  Src\,I and SMA1,
have been identified using subarcsecond angular resolution  and sensitive
ALMA data at both bands 6 and 7 \citep{Beuther2004,Hirota2015}.
One of the compact sources detected in the western region of the compact ridge
is associated with the  location of the strong occasionally  bursting water
masers \citep{Hirota2011,Hirta2014,Hirota2015}. No compact source
is detected in the northwestern clump where the continuum emission  detected in our data is completely
resolved in higher angular resolution ALMA data \citep{Hirota2015}.
In addition, three of the compact sources identified by \citet{Hirota2015} are not
located in one of the main molecular components. Two of these compact objects are
associated with infrared sources (IRc4 and IRc7)  while the remaining one is not
detected in our data due to  limited sensitivity \citep{Shuping2004,Hirota2015}.

\subsection{Spatial distribution of H$_2$CS}
In addition to the 1.2\,mm continuum, Fig.\,\ref{fig1} also shows the integrated intensity maps
of para-H$_2$CS\,$7_{07}-6_{06}$, $7_{25}-6_{24}$, and $7_{26}-6_{25}$ overlaid on the continuum emission.
It shows that the morphology of the H$_2$CS triplet is consistent with the continuum
emission, indicating that dense gas traced by para-H$_2$CS\,7--6 is closely  associated  with the
dust traced at 1.2\,mm in the Orion\,KL region. Few H$_2$CS emission peaks reveal minor offsets from
the continuum emission peaks (see Fig.\,\ref{fig1}). The largest offset is
$\sim$1\hbox{$\,.\!\!^{\prime\prime}$}8 ($\sim$750\,AU) between the peaks of continuum emission
and H$_2$CS in the elongated ridge.

The morphology of para-H$_2$CS\,7--6 emission in the Orion\,KL region is similar to results previously
obtained with other molecular tracers, e.g., NH$_3$, H$_2$CO, CH$_3$OH, CH$_3$CN, HCOOCH$_3$, HC$_3$N, and SO$_2$
(e.g., \citealt{Wilson2000,deVicente2002,Wang2010,Favre2011,Goddi2011,Gong2015,Peng2017HC3N,Luo2019,Pagani2019III}).
However, the northwestern clump was not always detected. For example, it was seen in the metastable
NH$_3$\,($J$,$K$)=\,(6,6)--(9,9) transitions but not in  transitions beyond $J$=$K$=\,9 \citep{Goddi2011}.
Furthermore, it was only detected in one  transition of the CH$_3$CN $J$=\,18-17 multi-line study by
\citet{Wang2010}, in the $K$=\,4 transition.

\subsection{H$_2$CS line ratios}
\label{line_ratios}
Since para-H$_2$CS\,$7_{25}-6_{24}$ and $7_{26}-6_{25}$ have similar upper state energies
above the ground state (see Sect.\,\ref{Data}) and similar line profiles
(for flux densities, line widths, and velocities of our data, see Fig.\,\ref{fig-spect}),
the averaged ratio para-H$_2$CS\,0.5$\times[(7_{25}-6_{24}+7_{26}-6_{25})/7_{07}-6_{06}$]
between para-H$_2$CS\,$7_{25}-6_{24}/7_{07}-6_{06}$ and $7_{26}-6_{25}/7_{07}-6_{06}$
is used in this work. Velocity-integrated intensity ratio maps of
para-H$_2$CS\,0.5$\times[(7_{25}-6_{24}+7_{26}-6_{25})/7_{07}-6_{06}$] detected above a signal-to-noise
ratio of $\sim$30 and derived from velocity integrated  intensities  in the Orion\,KL region are
shown in Fig.\,\ref{Fig2}. The para-H$_2$CS line ratios
range from 0.30 to 1.42 with an average of 0.68\,$\pm$\,0.16 (errors given here
and elsewhere are standard deviations of the mean) for  the colored regions in Fig.\,\ref{Fig2}.
Ratios are found between 0.65 and 1.42 with an average of 0.87\,$\pm$\,0.13
in the elongated ridge, between 0.38 and 0.99 with an average of
0.65\,$\pm$\,0.12 in the compact ridge, and between 0.30 and 0.91 with an average of
0.58\,$\pm$\,0.11 in the northwestern clump. Averages were calculated, taking each line ratio
value with the same weight. Fig.\,\ref{Fig2} shows immediately, even without a detailed analysis,
that ratios are particularly high near Src\,I and the supposed location of the
explosive event  to become lower farther away from it.  Locally, higher  para-H$_2$CS line ratios
are also associated with the western dense gas component of the compact ridge.

For the para-H$_2$CS triplet, compared to the single-dish APEX\,12-m data
(beam size $\sim$26$''$, \citealt{Brinkmann2020}),
the fluxes recovered by ALMA data are $\sim$70\%, $\sim$90\%, and $\sim$90\%
for para-H$_2$CS $7_{07}-6_{06}$, $7_{25}-6_{24}$, and $7_{26}-6_{25}$, respectively.
Since the maximum recoverable scale is $\sim$8\hbox{$\,.\!\!^{\prime\prime}$}7,
the missing flux may mostly be due to large scale structure filtered out by the interferometer,
and the larger amounts of the missing flux in the $7_{07}-6_{06}$ transition may be caused
by the fact that this transition requires less excitation to become detectable than the other two.
Thus line ratios  can be overestimated by not more than 22\% if the missing flux densities
are assumed to be $\sim$30\%, $\sim$10\%, and $\sim$10\% for para-H$_2$CS $7_{07}-6_{06}$,
$7_{25}-6_{24}$, and $7_{26}-6_{25}$, respectively. However, the more compact and dense the
regions are, the smaller the  proportion of the flux that  will be missed.
The para-H$_2$CS triplet traces compact emission regions,
and exhibits nearly identical structures in Orion\,KL (see Fig.\,\ref{fig1}), so the line ratios
of para-H$_2$CS may only be weakly influenced by filtered out emission.

\begin{figure*}[t]
\centering
\epsscale{1.15}
\plotone{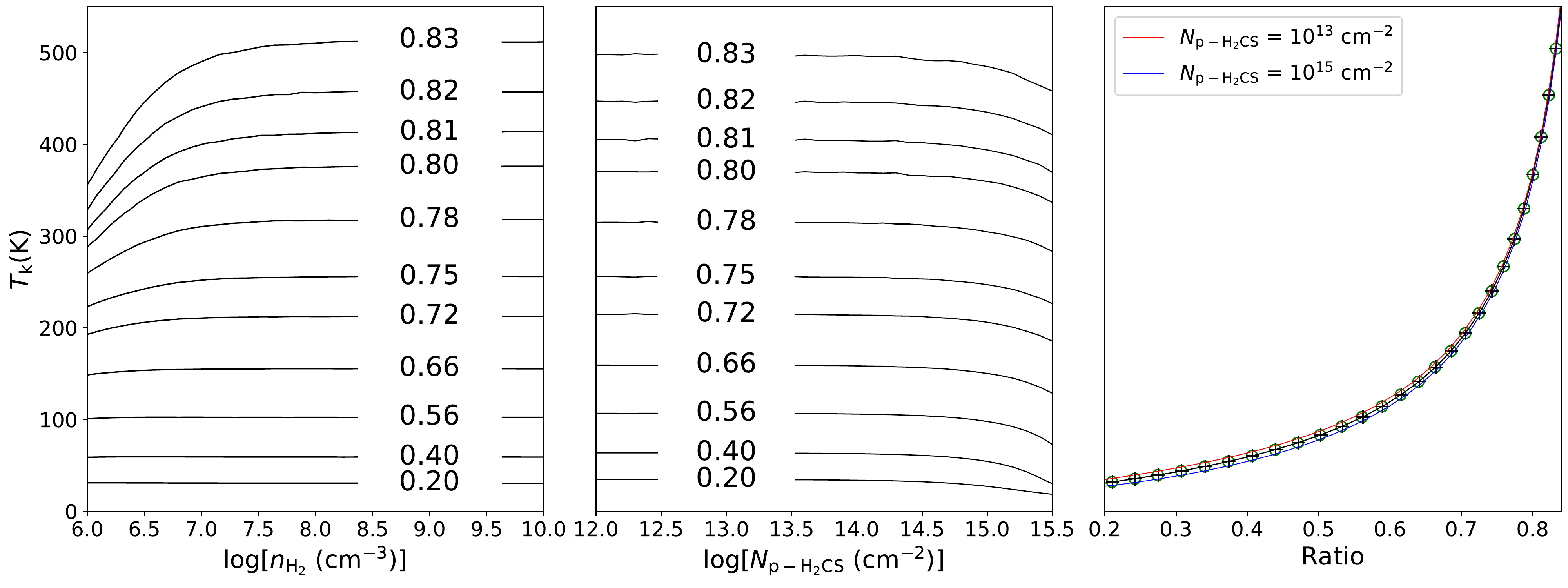}
\caption{Gas kinetic temperature modeled with RADEX, with the $T_{\rm {kin}}$ scale
on the left representing the ordinate for all three panels. The gas kinetic temperature
as a function of $n_{\rm H_2}$ ($N_{\rm p-H_2CS}=5\times10^{14}$\,cm$^{-2}$
and $\Delta v=5\,{\rm km\,s}^{-1}$) and para-H$_2$CS column density ($n_{\rm H_2}=10^{7}$\,cm$^{-3}$ and
$\Delta v=5\,{\rm km\,s}^{-1}$) are shown in the left and central panel, respectively.
Kinetic temperatures as a function of para-H$_2$CS line intensity ratios (the black line
stands for the above mentioned $N_{\rm p-H_2CS}=5\times10^{14}$\,cm$^{-2}$, $n_{\rm H_2}=10^{7}$\,cm$^{-3}$
and $\Delta v=5\,{\rm km\,s}^{-1}$) are shown in the right panel,
which demonstrates that line ratios are insensitive to any relevant parameter except kinetic temperature
as long as the lines are optically thin. Green circles provide para-H$_2$CS\,$7_{25}-6_{24}$/$7_{07}-6_{06}$
line ratios for $n_{\rm H_2}=10^{7}$\,cm$^{-3}$, $N_{\rm p-H_2CS}=5\times10^{14}$\,cm$^{-2}$,
and $\Delta v=5\,{\rm km\,s}^{-1}$, black plus symbols denote the corresponding values for
the $7_{26}-6_{25}$/$7_{07}-6_{06}$ line ratios. For para-H$_2$CS column densities of $10^{13}$
and $10^{15}$\,cm$^{-2}$ and the above mentioned density and line width,
para-H$_2$CS\,0.5$\times[(7_{25}-6_{24}+7_{26}-6_{25})/7_{07}-6_{06}$]
line ratios are plotted by red and blue solid lines, respectively. Note that red, blue, and black lines
become indistinguishable at high kinetic temperatures, even after magnification, on the scale of this plot.}
\label{redex_fig}
\end{figure*}

\subsection{Kinetic temperatures from H$_2$CS line ratios}
\label{H2CS_thermometer}
Using the non-local thermodynamic equilibrium (non-LTE) radiation transfer code
RADEX\footnote{https://home.strw.leidenuniv.nl/$\sim$moldata/radex.html}
\citep{radex2007} coupled with collision rates from the  Leiden Atomic and Molecular
Database\footnote{https://home.strw.leidenuniv.nl/$\sim$moldata/}
derived by \citet{Wiesenfeld2013}, a set of simulations has been carried out in order to explore
the relationship between the gas kinetic temperature
and the integrated intensity ratios of para-H$_2$CS\,$7_{25}-6_{24}$/$7_{07}-6_{06}$
and $7_{26}-6_{25}$/$7_{07}-6_{06}$ (see Fig.\,\ref{redex_fig}).
In the Orion\,KL region, previous measurements indicate that gas volume densities of H$_2$
are quite high, around $10^{7}$\,cm$^{-3}$, and even higher in compact regions
(e.g., \citealt{Gong2015,Pagani2017I}). Additionally, the column densities of H$_2$ measured
from H$_2$CO are $\gtrsim10^{24}$\,cm$^{-2}$ \citep{Mangum1993}.
Fig.\,\ref{redex_fig} shows that the gas kinetic temperatures estimated from
the para-H$_2$CS\,7--6 line ratios are nearly independent of the
H$_2$ volume density ($n_{\rm H_2}$) and para-H$_2$CS column density ($N_{\rm p-H_2CS}$) at
$n_{\rm H_2}\,\gtrsim\,10^{7}$\,cm$^{-3}$ and $N_{\rm p-H_2CS}\,\lesssim\,6\times 10^{14}$\,cm$^{-2}$ per km\,s$^{-1}$.
Therefore, the gas kinetic temperatures can be directly estimated from the integrated intensity
ratios of para-H$_2$CS.

Observations of H$_2$CS and its isotopologues suggest that para-H$_2$CS\,(7--6)
lines are optically thin in the Orion\,KL region \citep{Tercero2010,Luo2019}.
The total velocity integrated beam averaged ($\sim$\,10$^{\prime\prime}$)
column density of para-H$_2$CS is $\sim5\times10^{14}$\,cm$^{-2}$ obtained from
IRAM\,30-m observations \citep{Tercero2010}. Under the assumption of
optically thin conditions for the para-H$_2$CS\,(7--6) triplet,
$n_{\rm H_2}=10^{7}$\,cm$^{-3}$, and $N_{\rm p-H_2CS}=5\times 10^{14}$\,cm$^{-2}$ and an
adopted line width $\Delta v$\,=\,5\,km\,s$^{-1}$ from our data and a 2.73\,K background
temperature, the gas kinetic temperature is estimated from the line ratios of
para-H$_2$CS\,0.5$\times[(7_{25}-6_{24}+7_{26}-6_{25})/7_{07}-6_{06}$]
(see the right panel of Fig.\,\ref{redex_fig}).

As mentioned in Sect.\,\ref{Data}, para-H$_2$CS\,$7_{25}-6_{24}$ and $7_{26}-6_{25}$
have almost identical upper state energies above the ground state ($E_{\rm u}/k\,=\,98.8$\,K)
and, according to Sect.\,\ref{line_ratios}, similar observed line profiles.
The resulting gas kinetic temperatures derived from para-H$_2$CS\,$7_{25}-6_{24}$/$7_{07}-6_{06}$
and $7_{26}-6_{25}$/$7_{07}-6_{06}$ are similar  at $n_{\rm H_2} = 10^{7}$\,cm$^{-3}$
and $N_{\rm p-H_2CS} = 5\times 10^{14}$\,cm$^{-2}$ (see Fig.\,\ref{redex_fig}),
so we use the averaged integrated ratio of
para-H$_2$CS\,0.5$\times[(7_{25}-6_{24}+7_{26}-6_{25})/7_{07}-6_{06}$],
already introduced in Sect.\,\ref{line_ratios},  as a measure of excitation
to determine the gas temperature in this work. The simulated results indicate that
the optical depth for para-H$_2$CS\,7--6 decreases as the gas temperature increases.
The optical depths for $N_{\rm p-H_2CS}=5\times10^{14}$\,cm$^{-2}$
are less than 0.3 for para-H$_2$CS\,$7_{07}-6_{06}$ and less than 0.09
for para-H$_2$CS\,$7_{25}-6_{24}$ and $7_{26}-6_{25}$ when the gas temperature  is greater
than 43\,K. The optical depth for para-H$_2$CS\,$7_{07}-6_{06}$ decreases rapidly to 0.09
when the gas temperature increases to 100\,K.

Uncertainties of the gas temperature can be estimated from formula
($\mathrm{d}T_{\rm k}/\mathrm{d}{R})\times 1/(S/N)\times\sqrt{0.5+{R}^{2}}$,
where the $\mathrm{d}T_{\rm k}/\mathrm{d}{ R}$ represents  the first derivative of $T_{\rm k}$ as a
function of the $\it R$, which can be  calculated numerically.  $R$ represents the velocity-integrated
ratio of para-H$_2$CS\,0.5$\times[(7_{25}-6_{24}+7_{26}-6_{25})/7_{07}-6_{06}$] and  $S/N$ represents the
signal-to-noise ratio of the para-H$_2$CS\,$7_{07}-6_{06}$ line. For our data, the uncertainties
are \textless15\%, $\lesssim$20\%, and $\textgreater$25\% for gas temperatures $\lesssim$200, $\lesssim$400,
and $\textgreater$500\,K, respectively. The uncertainties estimated here are exclusively based on
the errors of the integrated line intensity fits, because the relative calibration errors
for the three lines should be very small in view of the simultaneous measurements and virtually
identical atmospheric conditions.

Generally, higher para-H$_2$CS line ratios indicate higher kinetic temperatures
(see Fig.\,\ref{redex_fig}, right panel), so the line ratio map of
para-H$_2$CS (Fig.\,\ref{Fig2}) can be used as a proxy
for  kinetic temperatures. The kinetic temperatures of dense gas derived from
the para-H$_2$CS\,7--6 line ratios in the Orion\,KL region are high and cover a wide range,
from 43 to $>$500\,K with an average of $\sim$170\,K at density $n_{\rm H_2}=10^{7}$\,cm$^{-3}$.
Gas kinetic temperatures  from 148 to $>$500\,K with an average of $>$500\,K are found
in the  elongated ridge, from 56 to $>$500\,K with an average
of $\sim$150\,K in the compact ridge, and from 43 to $>$500\,K with an average of $\sim$110\,K
in the northwestern clump.  The average values of gas kinetic temperatures are directly derived
from the averaged values  of the para-H$_2$CS\,(7--6) line ratios
(taking each line ratio  value with the same weight) presented in Sect.\,\ref{line_ratios},
making use of the  monotonic rise of $T_{\rm kin}$ with increasing integrated line intensity ratio.

\section{Discussion}
\label{Discussion}
\subsection{Gas temperatures}
\label{Comparison}
The gas kinetic temperatures obtained from para-H$_2$CS line ratios indicate a highly
non-uniform temperature distribution in Orion\,KL, which reveals that highest temperatures
($T_{\rm kin}$\,$>$\,500\,K) are associated with the elongated ridge region
and the location of the cluster of H$_2$O  masers  in the western compact ridge \citep{Hirota2011,Hirta2014}.
More  moderate temperatures (150\,$<$\,$T_{\rm kin}$\,$<$\,500\,K) are obtained for the
southeastern region of the northwestern clump and the eastern region of the compact ridge,
while lower temperatures ($T_{\rm kin}$\,$<$\,150\,K) characterize the northwestern region
of the northwestern clump and the southern region of the compact ridge.

A non-uniform temperature distribution in Orion\,KL was already partially revealed by previous observations.
For example, gas temperatures ranging from 130 to 170\,K in the elongated ridge and about $\sim$130\,K
near the compact ridge have been obtained from NH$_3$\,(4,4) and (10,9) observed with the VLA
(beam size\,$\sim$\,1$''$, see Fig.\,4 in \citealt{Wilson2000}). Gas temperatures varying between
160 and 490\,K in the elongated ridge and the compact ridge have been revealed by performing
NH$_3$ (6,6)--(12,12) observations made with the VLA (beam size\,$\sim$\,1$''$, see Fig.\,3
in \citealt{Goddi2011}). Gas rotation temperatures, $T_{\rm {rot}}$, ranging from 68 to 176\,K
have been calculated by SO$_2$ \citep{Luo2019}. The highest rotation temperatures ($\sim$176\,K) are
located in the elongated ridge and lower temperatures (68--106\,K) characterize the compact ridge
and the northwestern clump (see Fig.\,10 in \citealt{Luo2019}).
In addition to the mapping results, the rotation temperatures are 190--620\,K in the elongated ridge
and 170--280\,K in the compact ridge, estimated by performing LTE excitation analysis using a multi-line study
of CH$_3$CN\,18$_K$--17$_K$ toward the peak positions of a velocity-integrated intensity map \citep{Wang2010}.
The rotation temperatures are 93--321\,K in the elongated ridge and 88--186\,K near the compact
ridge measured by HC$_3$N at the peak positions of an H$^{13}$CCCN\,$\nu$\,=\,0 velocity-integrated
intensity map \citep{Peng2017HC3N}. Rotation temperatures are 103--140\,K in the elongated ridge,
119--168\,K in the compact ridge, and 108--111\,K in the northwestern region estimated by
HCOOCH$_3$ towards the peak positions of an HCOOCH$_3$ velocity-integrated intensity map \citep{Favre2011}.

Here we present not rotation but kinetic temperatures, making use of the suitable
properties of the H$_2$CS molecule. Furthermore, this molecule provides an
insight into the densest parts of Orion\,KL, revealing an unprecedented particularly
detailed view of the kinetic temperature and its gradients across the source.
Our resulting $T_{\rm kin}$ values reach maxima  near the supposed location of
the explosive event and become lower farther away from it (see Fig.\,\ref{Fig2}).
This has not been seen in such clarity before, e.g. in the rotation temperature maps derived
from NH$_3$ and SO$_2$ \citep{Wilson2000,Goddi2011,Luo2019}.
While NH$_3$ is thought to be an excellent and commonly used gas kinetic temperature
tracer of star forming regions \citep{Ho1983,Li2013,Lu2014,Friesen2017,Wu2018,Zhou2020},
it is a lower density gas tracer and is not  exclusively probing the very dense gas with
$\sim$10$^{7}$\,cm$^{-3}$ and even higher values in the Orion\,KL region.  For example,
the critical density\footnote{The critical density (assuming optically thin emission)
is estimated from $n_{\rm crit}\,=\,A_{ij}/C_{ij}$.
$A_{ij}$ and $C_{ij}$ represent the Einstein coefficient for spontaneous emission and the
collision rate, respectively. The values are obtained from  the  Leiden Atomic and Molecular
Database via https://home.strw.leidenuniv.nl/$\sim$moldata/.} of NH$_{3}$\,(6,6) is around
1.7--2.5\,$\times$\,10$^{3}$\,cm$^{-3}$, while the corresponding value for our para-H$_{2}$CS\,7--6
lines is $\sim$2\,$\times$\,10$^{6}$\,cm$^{-3}$  at kinetic temperatures 10--300\,K.
Besides, NH$_3$\,(6,6), where the hyperfine satellite lines were not covered, is probably
affected by line saturation. For example, we made an estimation of NH$_3$\,(6,6) opacity
$\tau_{(\rm NH_{3}(6,6))}$\,\textgreater\,1 by running RADEX under assumptions of
$T_{\rm {kin}}$\,=\,300\,K, $n_{\rm H_{2}}$\,=\,10$^{4}$\,cm$^{-3}$,
$N_{\rm ortho-NH_{3}}$\,=\,1.5$\times$10$^{17}$\,cm$^{-2}$, and $\delta v$\,=\,10\,km s$^{-1}$
following the results obtained from \citet{Goddi2011}. It needs to be confirmed through further
hyperfine structure observations (from a 40$^{\prime\prime}$ beam size single
dish measurement \citet{Hermsen1988} derive an overall  opacity of 14 (their Table 4)
for the (6,6) line in the hot core).

SO$_2$ is generally not thought to be a gas kinetic
temperature tracer and the rotation temperatures derived may differ from
the gas kinetic temperatures. In the cases of CH$_3$CN, HC$_3$N, and HCOOCH$_3$,
there are no rotation temperature maps available for the Orion\,KL region.
Based on present calculated results on several positions, low transitions of
CH$_3$CN\,(18$_K$-17$_K$;\,$K$\,=\,0--4) are optically thick in  Orion\,KL \citep{Wang2010}.
Furthermore, HCOOCH$_3$ is likely to trace lower temperatures of gas compared to other
molecular tracers \citep{Favre2011}. As mentioned in Sect.\ref{intro},
para-H$_2$CO\,($J$\,=\,3--2 and 4--3) is a good dense gas kinetic temperature tracer.
However, line ratios of para-H$_2$CO\,$J$\,=\,3--2 and 4--3 are insensitive to detect
temperatures in excess of 150\,K \citep{Mangum1993,Immer2016,Tang2018b}.

To summarize, for the specific case of the  very dense and hot Orion\,KL region
the molecular species mentioned above appear not to be suitable
to trace the kinetic temperature of the dense gas as well as H$_2$CS. Various limitations,
e.g. opacity, sensitivity, and poorly constrained correlations between excitation and kinetic
temperatures will affect the results. So, para-H$_2$CS appears to be quite a suitable
tracer to measure high temperatures of dense gas in Orion\,KL region.

\subsection{Impact of the explosive outflow}
\label{outflow}
High temperatures are associated with the northern and southwestern edge of the elongated
ridge region (see Fig.\,\ref{Fig2}). These two regions appear to be associated with the
outflow traced by SiO emission along a northeast-southwest axis emanating from
Src\,I (see Fig.\,1 in \citealt{Plambeck2009OUTFLOW}). However, this outflow is quite slow
($<$20\,km\,s$^{-1}$),  suggesting that the outflow may not play a significant role in heating the
dense gas in the northern edge and southwestern corner of the elongated
ridge region. Instead, the maximum velocity of the CO\,$J$\,=\,2--1 bullets produced from the explosion
can reach up to $\sim$160\,km\,s$^{-1}$ within a radius of $\sim$50$^{\prime\prime}$ ($\sim$0.1\,pc)
centred on the location of the  explosive event revealed by recent ALMA observations
with a beam size of \,$\sim$\,1$''$ \citep{Bally2017}. This indicates that it is the explosion,
which  strongly impacts on the dense gas in the $\sim$20$^{\prime\prime}$ sized region of Orion\,KL.
High velocity ($\textgreater$100\,km\,s$^{-1}$) CO\,$J$\,=\,2--1 and 3--2 bullets ejected
from the explosion and detected by the SMA appear to be blocked
and absorbed in the elongated ridge \citep{Zapata2009,Zapata2011}.
Further observations show that the dense regions including the compact ridge and the northwestern
clump also lack the CO bullets while strong class I CH$_3$OH maser emission in the $J_2-J_1~E$ transitions
near 25\,GHz $J = 4, 5, ...$ is found in this general area \citep{Johnston1992}. Class I CH$_3$OH masers
are excited by shocks \citep{Leurini2016}. However, there appears to exist no  direct spatial correspondence
between these masers and other energetic phenomena, a fact that remains to be investigated.

Previous observations of several molecular species, e.g., SiO, H$_2$CO, and CH$_3$OH,
show the filamentary structure of the Orion\,KL region \citep{Plambeck2009OUTFLOW,Gong2015,Pagani2019III},
which may be shaped by the explosion. To define the exact directions of the strong interaction
between explosive outflow and dense gas, high sensitivity ALMA observations of several molecular
species have been carried out \citep{Pagani2019III}. Elongated molecular tail structures along
the arrows 1--8 in Fig.\,\ref{Fig2} are clearly detected in H$_2$CO\,($9_{18}–9_{19}$),
which are interpreted in terms of a strong interaction of the dense gas with the high velocity
ejecta from the explosion \citep{Pagani2019III}. Several of these molecular tail structures are
also identified by our para-H$_2$CS\,($7_{07}-6_{06}$) data (see the arrows 4--8 in Fig.\,\ref{Fig2}),
but they are not as long as those seen in H$_2$CO \citep{Pagani2019III}, which can be used to trace
less dense gas.

\begin{figure}[t]
\centering
\epsscale{1.25}
\plotone{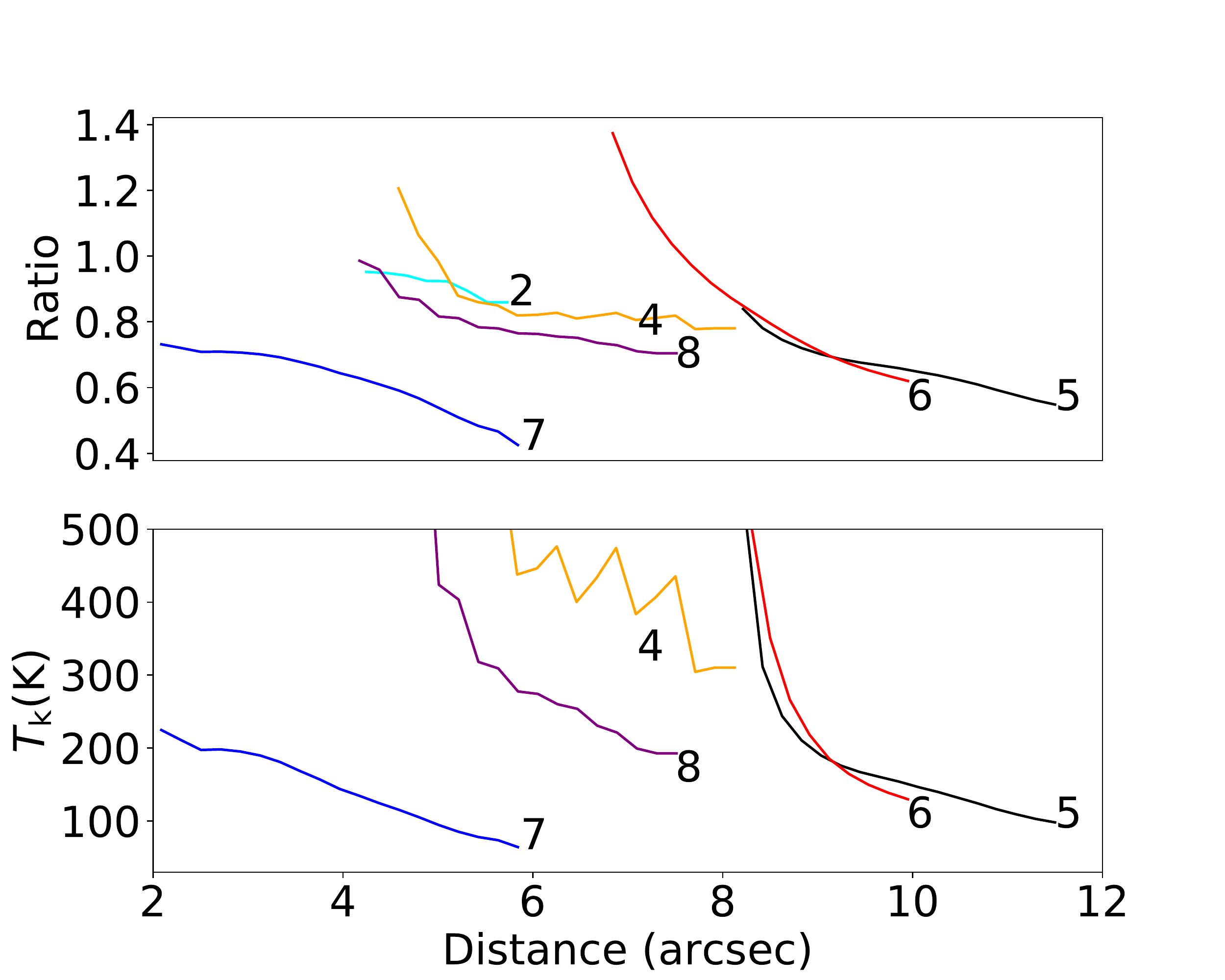}
\caption{Profiles of velocity-integrated intensity ratios of para-H$_2$CS (\emph{upper})
and gas kinetic temperatures (\emph{lower}) along the arrows from the explosive
center to the dense gas shown in Fig.\,\ref{Fig2}. The intensity ratio is the one also
displayed in Fig.\,\ref{Fig2}. Because $T_{\rm kin}$\,$>$\,500\,K values are not as
reliable as those $<$\,500\,K (Sect.\,\ref{H2CS_thermometer}), we have cut the ordinate
of the lower panel at 500\,K, which implies that there data from arrow 2 are missing.
Different colors of lines correspond to different molecular tails along the arrows
shown in Fig.\,\ref{Fig2}.}
\label{lines_fig}
\end{figure}

Following the directions of the molecular tails shown in Fig.\,\ref{Fig2}, we investigate the para-H$_2$CS
line ratios and gas temperature gradients in Fig.\,\ref{lines_fig}. It shows significant gradients of
para-H$_2$CS line ratios and gas kinetic temperatures along these molecular tails.
The farther from the location of the explosive event a
molecular tail is starting, the higher the ratios  and the gas kinetic temperature
at the closest approach is (see Figs.\,\ref{Fig2} and \ref{lines_fig}).
This might be a consequence of our viewing angle
towards the source, providing projected but not real distances. The temperature gradients
along the molecular tails 2, 4, 5, 6, 7, and 8 identified
with para-H$_2$CS (see Figs.\,\ref{Fig2} and
\ref{lines_fig}) indicate that the dense gas is heated by shocks initiated by the explosion,
which occurred around 550 yr ago. This was not revealed by previous gas temperature measurements
with NH$_3$ (4,4) and(10,9), and (6,6)--(12,12) \citep{Wilson2000,Goddi2011} or SO$_2$ \citep{Luo2019}
and provides direct evidence for most parts of the Orion\,KL region being heated by an external source.
This agrees with previous suggestions
(e.g., \citealt{Wilson2000,Wang2010,Wang2011,Goddi2011,Favre2011,Zapata2011,Orozco2017,Peng2017HC3N}).

Very high temperatures are not associated with the dense gas in the northwestern clump (see Fig.\,\ref{Fig2}).
However, this clump also shows a distinct temperature gradient  (see the molecular tail 7
in Figs.\,\ref{Fig2} and \ref{lines_fig}). This suggests that the dense gas in the northwestern
clump is also most likely heated by shocks from the explosive outflow. The generally lower $T_{\rm kin}$
values may be due to a location either a little in front or behind the elongated ridge.
Additionally, the motions of the CO high velocity ejecta towards the northwest might follow
the plane of the sky \citep{Bally2017}, so that only a small percentage of the ejecta have interacted
with the northwestern clump, leading to lower temperatures than in the other regions.

High temperatures are found inside a tiny region near the southeastern edge of the northwestern clump.
This region is associated with continuum source IRc7 (see Fig.\,\ref{Fig2}), suggesting that the continuum source
IRc7 has some impact on its immediate neighbourhood. However, because of its spatially limited impact,
we still find that the  dense gas in the northwestern clump is predominantly heated by the shocks produced
from the explosion.

In the compact ridge, which is in projection farther away from the site of the explosion,
internal heating plays a more important role. Intense 22 GHz H$_2$O maser emission
\citep{Hirota2011,Hirta2014} is associated with the western half of the compact ridge
(see Fig.\,\ref{Fig2}). This indicates that here the high temperature
of the dense gas may be dominated by local massive star formation activity.

\section{Conclusions}
\label{Conclusions}
We have mapped the kinetic temperature structure of Orion\,KL in a $\sim$20$''$ ($\sim$8000\,AU)
sized region making use of the para-H$_2$CS $7_{07}-6_{06}$, $7_{25}-6_{24}$, and $7_{26}-6_{25}$
lines in the ALMA Band 6 at frequencies near 240\,GHz. The main results are summarised as follows.

\begin{enumerate}
\item
The kinetic temperatures are derived from the measured averaged velocity-integrated
intensity ratios of para-H$_2$CS\,$7_{26}-6_{25}/7_{07}-6_{06}$ and $7_{25}-6_{24}/7_{07}-6_{06}$
with RADEX non-LTE modeling, which provides a monotonic
increase in line ratios with rising kinetic temperatures.
Kinetic temperatures range from 43 to $>$500\,K with an unweighted average of  $\sim$170\,K
at a spatial density of 10$^7$\,cm$^{-3}$.

\item
Our measurements of dense gas temperatures from para-H$_2$CS\,$J$\,=\,7--6 line ratios
reveal a temperature gradient from the explosive center to  the outer regions of Orion\,KL.
Significant gradients of kinetic temperature along molecular filaments traced by H$_2$CS
indicate that the dense gas is heated by the shocks induced by the enigmatic
event that occurred 550\,yr ago. Internal sources do not play a dominant role in heating the
elongated ridge, the northwestern clump, and the eastern region of the compact ridge,
while the higher temperatures in the western  compact ridge are likely caused
by internal massive star formation activity. Furthermore, the kinetic temperature may
also be enhanced by IRc4 and IRc7, but only in tiny, spatially highly confined regions.

\end{enumerate}

\acknowledgments
We thank the anonymous referee for improving the paper.
This work was funded by the CAS “Light of West China” Program under grant 2016-QNXZ-B-23,
the National Natural Science foundation of China under grants 11703074, 11903070, 11433008,
11703073, and 11973076, the “TianShan Youth Plan” under grant 2018Q084, the Heaven Lake
Hundred-Talent Program of Xinjiang Uygur Autonomous Region of China, and the Collaborative Research
Council 956, subproject A6, funded by the Deutsche Forschungsgemeinschaft (DFG).
C.\,H. has been funded by Chinese Academy of Sciences President's International
Fellowship Initiative with grant No.\,2021VMA0009. G.\,W. acknowledges the support from
Youth Innovation Promotion Association CAS. This paper makes use of the
following ALMA data: ADS/JAO.ALMA\#2011.0.00009.SV. ALMA is a partnership
of ESO (representing its member states), NSF (USA) and NINS (Japan),
together with NRC (Canada), MOST and ASIAA (Taiwan), and KASI (Republic of Korea),
in cooperation with the Republic of Chile. The Joint ALMA Observatory is operated by ESO,
AUI/NRAO and NAOJ. This research has used NASA's Astrophysical Data System (ADS).

\bibliographystyle{aasjournal} 
\bibliography{ref} 

\end{document}